\documentstyle[12pt]{article}
\begin{document}
\tolerance=5000
\def\be{\begin{equation}}
\def\ee{\end{equation}}
\def\bea{\begin{eqnarray}}
\def\eea{\end{eqnarray}}
\def\beaa{\begin{eqnarray*}}
\def\eeaa{\end{eqnarray*}}
\def\nn{\nonumber \\}
\def\cF{{\cal F}}
\def\det{{\rm det\,}}
\def\Tr{{\rm Tr\,}}
\def\e{{\rm e}}
\def\etal{{\it et al.}}
\def\erp2{{\rm e}^{2\rho}}
\def\erm2{{\rm e}^{-2\rho}}
\def\er4{{\rm e}^{4\rho}}
\def\etal{{\it et al.}}

\ 

\vskip -2cm

\ \hfill
\begin{minipage}{3.5cm}
November 1999 \\
\end{minipage}

\vfill

\begin{center}
{\Large\bf Finite gravitational action for higher 
derivative and stringy gravities}

\vfill

{\sc Shin'ichi NOJIRI}\footnote{\scriptsize 
e-mail: nojiri@cc.nda.ac.jp, snojiri@yukawa.kyoto-u.ac.jp} and
{\sc Sergei D. ODINTSOV$^{\spadesuit}$}\footnote{\scriptsize 
e-mail: odintsov@mail.tomsknet.ru, odintsov@itp.uni-leipzig.de}

\vfill

{\sl Department of Mathematics and Physics \\
National Defence Academy, 
Hashirimizu Yokosuka 239, JAPAN}

\ 

{\sl $\spadesuit$ 
Tomsk Pedagogical University, 634041 Tomsk, RUSSIA \\
}

\ 

\vfill

{\bf abstract}
\end{center}

We generalize the local surface counterterm prescription suggested 
in Einstein gravity for higher derivative (HD) and Weyl gravities. 
Explicitly, the surface counterterm is found for three- 
and five-dimensional HD gravities. As a result, the gravitational 
action for asymptotically AdS spaces is finite and gravitational 
energy-momentum tensor is well-defined. 
The holographic trace anomaly for d2 and d4 boundary (gauge) QFT dual 
to above HD gravity is calculated from gravitational energy-momentum
tensor. The calculation of AdS black hole mass in HD gravity is 
presented within above prescrition. The comparison with the standard 
prescription (using reference spacetime) is done.

\newpage

\section{Introduction}

One of the important achievements in AdS/CFT correspondence 
(for a general review and list of references, see \cite{ASMOO}) is 
related with the possibility to make the the finite gravitational 
action and to find the quasilocal stress tensor in gravity 
using a local surface counterterm prescription \cite{2}. 
With such prescription one chooses the coordinate 
invariant functional of the intrinsic boundary geometry and adds 
it to gravitational action. As a result, the equations of motion 
are not modified. 
However, the specific choice of this functional (often called the 
surface counterterm) cancels the divergences of gravitational action 
on the background under consideration. Then, one comes to 
well-defined expression for gravitational energy-momentum tensor 
and action \cite{2}.

 From the very beginning, the local counterterm prescription 
looks more elegant than the standard prescription of subtraction 
of reference spacetime. There, the action is regulated by 
restriction of spacetime to the interior of some
boundary geometry and then, by subtraction of (infinite) 
gravitational action of another (reference) background 
which has the same boundary \cite{3}. Clearly, the standard 
prescription is ambiguous as it depends from the choice of 
reference spacetime. Moreover, it is not always possible
to select the necessary reference background. Very often 
the embedding of boundary geometry in the reference spacetime 
may be done only approximately.
Nevertheless, in many cases the standard prescription works 
and finite gravitational action is obtained in the limit of 
infinite boundary. The quasilocal stress tensor could be 
then found by variation of finite gravitational 
action over surface metric \cite{4}.  

The local surface counterterms prescription \cite{2} has been successfully 
tested in asymptotically AdS spacetimes. It reproduces correctly 
the mass and angular momentum for various AdS backgrounds. 
The results coincide with the ones for same background (say, 
energy evaluation \cite{6,7,8,9,10,11}) using standard procedure. 
Explicitly, the surface counterterms (finite polynomials) 
for $d\leq 4$, $d\leq 6$, $d\leq 8$ asymptotically AdS 
spacetimes have been found in refs.\cite{2,5,12} respectively.

It is very interesting that such prescription has clear 
similarity (via AdS/CFT correspondence) with usual counterterms 
prescription (renormalization) in quantum field theory. 
Moreover, as it was shown in ref.\cite{13}(see also ref.\cite{ho}) the local
surface counterterm prescription admits the natural continuation for
asymptotically flat spaces.

So far, the consideration of above prescription has been limited to
Einstein gravity where the universal and widely accepted 
Gibbons-Hawking boundary term \cite{3} is known. The role of 
this term is to make the variational procedure to be 
well-defined one. 

It would be extremely interesting to test the universality 
of prescription \cite{2} applying it to another gravitational 
theories. The purpose of this work is to formulate the local 
counterterm prescription for higher derivative (or $R^2$) 
gravity. As is known such theory is good alternative for 
Einstein gravity. Moreover, it could be that higher derivative 
(HD) terms are relevant only near Planck scale as there may be 
mechanisms to make such HD terms negligible at present stage 
of Universe evolution. Such theory is well studied as
perturbative QFT (for a review, see \cite{BOS}). HD gravity has 
better ultraviolet properties than Einstein one. From another side,
HD gravity often appears as low-energy effective action from 
string theory. Hence, it may be considered as stringy gravity.
The important difference of HD gravity from Einstein gravity is the fact that 
boundary term (analog of Gibbons-Hawking term) is not only much more
complicated but also there is no standard form of such term accepted in the
literature. The reason is that boundary term also includes the HD terms. 
Hence, it is quite difficult to make the variational procedure for HD gravity 
to be well-defined (unlike the case of Einstein gravity).

In the next section we discuss the boundary term in HD gravity and in its 
special version (Weyl or stringy) gravity. The local surface counterterm in 
such theories is explicitly found for d2 and d4 cases,
i.e. when the corresponding gravity theory 
(bulk) is formulated in three and five
dimensions.
 Comparison with the Einstein gravity is made. As a result the finite
gravitational action for HD gravity
 is found as well as gravitational energy-momentum tensor. From this 
as an explicit check we define the holographic conformal anomaly 
in d2 and d4 (via AdS/CFT correspondence). This conformal anomaly 
found from finite gravitational action in three and five dimensions coincides 
with the previous calculation of holographic conformal anomaly using other
methods.
In third section the calculation of AdS black hole mass is done for HD
gravity. In this calculation the local surface counterterm prescription is
used.
It is shown that the result coincides with the corresponding calculation with 
the help of standard prescription. Sometimes, there may appear some finite
difference between the calculation in two different prescriptions.
This difference is regarded as regularization dependence. We show 
how one can make two different results coinciding by fixing of some 
arbitrary parameters in local surface counterterm for HD gravity.
In last section the short summary is given.

\section{Surface counterterm and finite gravitational action}

In the present section we will define the local surface counterterm 
for HD gravity in three and five dimensions. This will make the
corresponding gravitational 
action for asymptotically AdS spaces to be finite.
We start with the action:
\be
\label{vi}
S=\int_{M_{d+1}} d^{d+1} x \sqrt{-\hat G}\left\{a \hat R^2 
+ b \hat R_{\mu\nu}\hat R^{\mu\nu}
+ c \hat R_{\mu\nu\xi\sigma}\hat R^{\mu\nu\xi\sigma}
+ {1 \over \kappa^2} \hat R - \Lambda \right\}\ .
\ee
Here $M_{d+1}$ is  $d+1$ dimensional manifold whose boundary is 
$d$ dimensional manifold $M_d$. The conventions of curvatures are 
given by
\bea
\label{curv}
R&=&g^{\mu\nu}R_{\mu\nu} \nn
R_{\mu\nu}&=& -\Gamma^\lambda_{\mu\lambda,\kappa}
+ \Gamma^\lambda_{\mu\kappa,\lambda}
- \Gamma^\eta_{\mu\lambda}\Gamma^\lambda_{\kappa\eta}
+ \Gamma^\eta_{\mu\kappa}\Gamma^\lambda_{\lambda\eta} \nn
\Gamma^\eta_{\mu\lambda}&=&{1 \over 2}g^{\eta\nu}\left(
g_{\mu\nu,\lambda} + g_{\lambda\nu,\mu} - g_{\mu\lambda,\nu} 
\right)\ .
\eea
When $a=b=c=0$, the action (\ref{vi}) becomes that of the 
Einstein gravity:
\be
\label{viE}
S=\int_{M_{d+1}} d^{d+1} x \sqrt{-\hat G}\left\{
{1 \over \kappa^2} \hat R - \Lambda \right\}\ .
\ee
If we choose 
\be
\label{viWa}
a={2 \over d(d-1)}\hat c \ ,\quad b= -{4 \over d-1}\hat c\ ,\quad
c=\hat c
\ee
the HD part of action is given by the square of the Weyl tensor 
$C_{\mu\nu\rho\sigma}$ : 
\be
\label{viWb}
S=\int_{M_{d+1}} d^{d+1} x \sqrt{-\hat G}\left\{
\hat c \hat C_{\mu\nu\xi\sigma}\hat C^{\mu\nu\xi\sigma}
+ {1 \over \kappa^2} \hat R - \Lambda \right\}\ .
\ee
The string theory dual to ${\cal N}=2$ superconformal field 
theory is presumably IIB string on ${\rm AdS}_5\times
X_5$ \cite{AFM} where $X_5=S^5/Z_2$. (The ${\cal N}=2$ $Sp(N)$ 
theory arises as the low-energy theory on the world volume on 
$N$ D3-branes sitting inside 8 D7-branes at an O7-brane). 
Then in the absence of Weyl term, ${1 \over \kappa^2}$ and $\Lambda$ are
given by 
\be
\label{prmtr}
{1 \over \kappa^2}={N^2 \over 4\pi^2}\ ,\quad
\Lambda= -{12N^2 \over 4\pi^2}\ .
\ee
This defines the bulk gravitational theory dual to suer YM theory 
with two supersymmetries.
The Riemann curvature squared term in the above bulk action 
may be deduced from heterotic string via heterotic-type I duality 
\cite{Ts2}. Using field redefinition ambiguity \cite{GWT} 
one can suppose that there exists the scheme where 
$R_{\mu\nu\alpha\beta}^2$ may be modified to 
$C_{\mu\nu\alpha\beta}^2$ in the same way as in ref.\cite{Marc}. 
Then, the action (\ref{viE}) is presumably the bulk action dual 
to ${\cal N}=2$ SCFT. 
In \cite{NOw}, for the ${\cal N}=2$ theory with the gauge group 
$Sp(N)$, the usual UV Weyl anomaly via AdS/CFT correspondence is reproduced if
\be
\label{W12}
\tilde c = {6N-1 \over 24\cdot 16\pi^2}\ ,\quad 
{1 \over \kappa^2}={12N^2 + 12 N -1 \over 3\cdot 16\pi^2}\ .
\ee
The second equation in (\ref{W12}) is, however, not compatible 
with (\ref{prmtr}), where ${1 \over \kappa^2}={N^2 \over 4\pi^2}$.  
This might suggest that some 
sub-leading corrections to ${1 \over \kappa^2}$ and/or $\Lambda$ 
would be necessary. The explicit form of such sub-leading terms has 
been defined in ref.\cite{NOw}. We call the model corresponding to 
(\ref{viWb}) as stringy gravity.

When $a$, $b$ and $c$ are given by
\be
\label{viGB}
a=c=\hat a \ , \quad b=-4\hat a
\ee
for $d=3$, the $R^2$ terms in (\ref{vi}) form  
the Gauss-Bonnet invariant $\tilde G$:
\be
\label{viGBb}
S=\int_{M_{d+1}} d^{d+1} x \sqrt{-\hat G}\left\{
\hat a \tilde G + {1 \over \kappa^2} \hat R - \Lambda \right\}\ .
\ee

The variation of the action (\ref{vi}) 
with respect to $\hat G^{\mu\nu}$ is given by
\bea
\label{Ii}
\delta S&=&\delta S_{M_{d+1}} + \delta S_{M_d} \\
\delta S_{M_{d+1}}&=&\int_{M_{d+1}} d^{d+1} x \sqrt{-\hat G}
\delta \hat G^{\zeta\xi}\left[
-{1 \over 2}G_{\zeta\xi}\left\{a \hat R^2 
+ b \hat R_{\mu\nu}\hat R^{\mu\nu} \right. \right. \nn
&& \left.+ c \hat R_{\mu\nu\xi\sigma}\hat R^{\mu\nu\xi\sigma}
+ {1 \over \kappa^2} \hat R - \Lambda \right\} \nn
&& \left. + 2a R R_{\zeta\xi} 
+ 2b \hat R_{\mu\zeta}{\hat R^{\mu}}_\xi
+ 2c \hat R_{\zeta\mu\nu\xi}\hat R_\xi^{\mu\nu\xi}
+ {1 \over \kappa^2} \hat R_{\zeta\xi}+ 
\cdots \right] \nn
\delta S_{M_d}&=& \int_{M_d} d^d x \sqrt{-\hat g}n_\mu\left[
2a\hat R\left\{\partial^\mu
\left(\hat G_{\xi\nu}\delta \hat G^{\xi\nu}\right) 
- D_\nu \left(\delta \hat G^{\mu\nu}\right)\right\} \right. \nn
&& + 2b\left\{{1 \over 2}\hat R^{\mu\nu}\partial_\nu
\left(\hat G_{\xi\sigma}\delta \hat G^{\xi\sigma}\right) 
-\hat R_{\nu\sigma}D^\sigma\left(\delta \hat G^{\mu\nu}\right) 
\right. \nn
&& \left. + {1 \over 2}\hat R_{\xi\sigma}D^\mu
\left(\delta \hat G^{\xi\sigma}\right)\right\}
+4c \hat R^{\mu\nu\xi\sigma} D_\xi
\left(\hat G_{\nu\tau}\hat G_{\sigma\kappa}
\delta \hat G^{\tau\kappa}\right) \nn
&& \left. + {1 \over \kappa^2}\left\{\partial^\mu
\left(\hat G_{\xi\nu}\delta \hat G^{\xi\nu}\right) 
- D_\nu \left(\delta \hat G^{\mu\nu}\right)\right\} \right]\ .
\nonumber
\eea
Here $\cdots$ in $\delta S_{M_{d+1}}$ expresses the terms 
containing the covariant derivative of the curvatures, 
$\hat g_{\mu\nu}$ is the metric on $M_d$ induced from 
$\hat G_{\mu\nu}$ and $n_\mu$ is the unit vector normal to 
$M_d$. 
We now choose the metric in the 
following form:
\be
\label{i}
ds^2\equiv\hat G_{\mu\nu}dx^\mu dx^\nu 
= {l^2 \over 4}\rho^{-2}d\rho d\rho + \sum_{i=1}^d
\hat g_{ij}dx^i dx^j \ , \quad 
\hat g_{ij}=\rho^{-1}g_{ij} \ .
\ee
Then $n^\mu$ and its covariant derivatives are given by,
\be
\label{Iiii}
n^\mu=\left({2\rho \over l},0,\cdots,0\right)\ ,\ \ 
D_\rho n^\rho = D_\rho n^i = D_i n^\rho=0\ ,\ \ 
D_in^j={\rho \over l}\hat g^{ik}\hat g'_{kj}\ .
\ee
Here $'\equiv {\partial \over \partial_\rho}$.
In the coordinate choice (\ref{i}), the surface terms 
$\delta S_{M_d}$ in (\ref{Ii}) have the 
following form
\bea
\label{Iii}
\delta S_{M_d}&=&\lim_{\rho\rightarrow 0}
\int_{M_d} d^d x \sqrt{-\hat g}{2\rho \over l}
\left[ 2 a
\hat R \left(\hat g_{ij}\delta \hat g^{ij}\right)' \right. \nn
&& + b\left\{\hat {R_\rho}^\rho\partial_\rho
\left(\hat g_{ij}\delta \hat g^{ij}\right) '
+ \hat R_{ij}\left(\delta \hat {g^{ij}}'
+ \hat g^{ik}\hat g_{kl} \delta \hat g^{lj}\right)\right\} \nn
&& \left. +4 c {\hat R^{\rho\ \rho}}_{\ i\ j} 
\left(\delta \hat {g^{ij}}'
+ \hat g^{ik}\hat g_{kl} \delta \hat g^{lj} \right) 
+ {1 \over \kappa^2}\partial^\rho
\left(\hat g_{ij}\delta \hat g^{ij}\right)  \right]\ .
\eea
Note that the terms containing $\delta\hat g^{\rho\rho}$ or 
$\delta\hat g^{\rho i}$ vanish. The variation $\delta S_{M_d}$ 
contains the derivative of $\delta \hat g^{ij}$ with respect 
to $\rho$, which makes the variational principle ill-defined. 
In order that the variational principle is well-defined on the 
boundary, the variation of the action should be written in the 
form of
\be
\label{bdry1}
\delta S_{M_d}=\lim_{\rho\rightarrow 0}
\int_{M_d} d^d x \sqrt{-\hat g}\delta\hat g^{ij}
\left\{\cdots\right\}
\ee
after using the partial integration. If we put 
$\left\{\cdots\right\}=0$ for $\left\{\cdots\right\}$ in 
(\ref{bdry1}), we could obtain the boundary condition. If 
the variation of the action on the boundary contains 
$(\delta \hat g^{ij})'$, however, we cannot partially integrate 
it with respect to $\rho$ on the boundary
to rewrite the variation in the form of (\ref{bdry1}) since 
$\rho$ is the coordinate expressing the direction perpendicular 
to the boundary. Therefore the ``minimum'' of the action is 
ambiguous. Such a problem was well studied in \cite{3} for the 
Einstein gravity ($a=b=c=0$) and the boundary term 
was added to the action, which cancels the variation : 
\be
\label{bdry2}
S_b^{\rm GH} = -{2 \over \tilde \kappa^2}
\int_{M_d} d^d x \sqrt{-\hat g}
 D_\mu n^\mu \ .
\ee
In the coordinate choice  (\ref{i}), the action (\ref{bdry2}) 
has the form 
\be
\label{bdry3}
S_b^{\rm GH} = -{2 \over \tilde \kappa^2}
\int_{M_d} d^d x \sqrt{-\hat g}
{\rho \over l}\hat g_{ij}\left(\hat g_{ij}\right)'\ .
\ee
Then the variation over the metric $\hat g_{ij}$ gives 
\be
\label{bdry4}
\delta S_b^{\rm GH} = -{2 \over \tilde \kappa}
\int_{M_d} d^d x \sqrt{-\hat g}
{\rho \over l}\left[ \delta \hat g^{ij}
\left\{-\hat g_{ik}\hat g_{il}\left(\hat g_{kl}\right)'
-{1 \over 2}\hat g_{ij}\hat g_{kl}\left(\hat g_{kl}\right)'
\right\}+ \hat g_{ij}\left(\delta\hat g_{ij}\right)'\right]\ .
\ee
 From the other side, the surface terms in the variation of 
the bulk Einstein action 
($a=b=c=0$ in (\ref{Iii})) have the form 
\be
\label{bdry5}
\delta S^{\rm Einstein}_{M_d}
=\lim_{\rho\rightarrow 0} {1 \over \kappa^2}
\int_{M_d} d^d x \sqrt{-\hat g}{2\rho \over l}
\left[ \hat g_{ij}' \delta \hat g^{ij} 
+ \hat g_{ij}\left(\delta \hat g^{ij}\right)'  \right]\ .
\ee
Then we find the terms containing 
$\left(\delta \hat g^{ij}\right)' $ in (\ref{bdry4}) and 
(\ref{bdry5}) are cancelled with each other. 

For HD gravity with the square of the scalar 
curvature and the square of the Weyl tensor 
$C_{\mu\nu\rho\sigma}$, 
where the action is given by
\be
\label{viW}
S=\int_{M_{d+1}} d^{d+1} x \sqrt{-\hat G}\left\{a \hat R^2 
+ \hat c \hat C_{\mu\nu\xi\sigma}\hat C^{\mu\nu\xi\sigma}
 \right\}\ ,
\ee
the boundary terms were proposed in \cite{HL} as follows:
\bea
\label{HL1}
S_b^{\rm HL} &=& \int_{M_d} d^d x \sqrt{-\hat g}\left[
4\left(a + {2\hat c \over d(d-1)}\right) \hat R D_\mu n^\mu 
\right. \nn
&& - {8\hat c \over d-1}\left(n_\mu n_\nu \hat R^{\mu\nu} 
D_\sigma n^\sigma + \hat R_{\mu\nu}D^\mu n^\nu \right) \nn
&& \left. 
+ 8\hat c n_\mu n_\nu \hat R^{\mu\tau\nu\sigma} 
D_\tau n_\sigma \right] \ .
\eea
For  general $R^2$ gravity (\ref{vi}), 
if the curvatures would not contain 
$\left(\hat g_{ij}\right)'$ nor 
$\left(\hat g_{ij}\right)''$, which appear in $\delta S_{M_d}$ 
of (\ref{Ii}), the boundary term would be given by
\bea
\label{HL2}
S_b^{R^2} &=& \int_{M_d} d^d x \sqrt{-\hat g}\left[
4 a\hat R D_\mu n^\mu 
+ 2 b\left(n_\mu n_\nu \hat R^{\mu\nu} 
D_\sigma n^\sigma + \hat R_{\mu\nu}D^\mu n^\nu \right) \right. \nn
&& \left. 
+ 8 c n_\mu n_\nu \hat R^{\mu\tau\nu\sigma} D_\tau n_\sigma 
- {2 \over \tilde \kappa^2}D_\mu n^\mu \right] \ .
\eea
If we choose $a \rightarrow a+{2 \over d(d-1)}\hat c$, 
$b \rightarrow -{4 \over d-1}\hat c$, $c\rightarrow \hat c$ 
(see Eq.(\ref{viWa})), the boundary terms for the action with 
the squares of the scalar 
curvature and the Weyl tensor correspond to the 
boundary action (\ref{HL1}) in \cite{HL}. 
In the coordinate choice (\ref{i}), $S_b^{R^2}$ in (\ref{HL2}) 
has the following form:
\bea
\label{HL3}
S_b^{R^2}&=&\lim_{\rho\rightarrow 0}
\int_{M_d} d^d x \sqrt{-\hat g}{\rho \over l}\left[
-4 a\hat R \hat g_{ij} \left(\hat g^{ij}\right)'
- 2 b\left({\hat R_\rho}^\rho \hat g_{ij} 
\left(\hat g^{ij}\right)' 
+ \hat R_{ij} \left(\hat g^{ij}\right)' \right) \right. \nn
&& \left. + 8 c {\hat R^{\rho}}_{i\rho j} 
\left(\hat g^{ij}\right)' - {2 \over \tilde\kappa^2}
\hat g_{ij} \left(\hat g^{ij}\right)'\right]\ .
\eea
In the coordinate choice (\ref{i}), the curvatures appearing in 
(\ref{HL3}) have the following forms:
\bea
\label{HL4}
R&=&\rho R + {3\rho^2 \over l^2}\hat g^{ij} \hat g^{kl}
\hat g_{ik}' \hat g_{jl}' - {4\rho^2 \over l^2}
\hat g^{ij} \hat g_{ij}'' 
- {\rho^2 \over l^2}\hat g^{ij} \hat g^{kl}
\hat g_{ij}' \hat g_{kl}' \nn
R_{\rho\rho}&=&{1 \over 4}\hat g^{ij} \hat g^{kl}
\hat g_{ik}' \hat g_{jl}' \nn
R_{ij}&=& R_{ij} + {2\rho^2 \over l^2} \hat g^{kl}
\hat g_{ik}' \hat g_{jl}' - {2\rho^2 \over l^2} \hat g_{ij}'' 
- {\rho^2 \over l^2} \hat g^{kl} \hat g_{ij}' \hat g_{kl}' \nn
R_{\rho i \rho j}&=&{1 \over 4} \hat g^{kl}
\hat g_{ik}' \hat g_{jl}' - {1 \over 2}\hat g_{ij}''\ .
\eea
Here $R$ and $R_{ij}$ are scalar and the Ricci curvatures 
given by $g_{ij}$ in (\ref{i}).  
We should note that the curvatures contain 
$\left(\hat g_{ij}\right)'$ and/or 
$\left(\hat g_{ij}\right)''$. Therefore the variation of 
the curvatures in the surface terms in (\ref{HL2}) or (\ref{HL3}) 
induces new terms containing $\left(\delta \hat g^{ij}\right)'$ 
and $\left(\delta \hat g^{ij}\right)''$. Therefore we need to add 
more surface terms to cancel them. It seems very difficult to 
 cancel the terms 
containing $\left(\delta \hat g^{ij}\right)''$. 
Instead of (\ref{HL2}), we choose $S_b$ by new parameters 
$\tilde a$, $\tilde b$, $\tilde c$ and $\tilde\kappa$ 
in the following form:
\bea
\label{Iiv}
S_b&=&S_b^{(1)} + S_b^{(2)} \nn
S_b^{(1)} &=& \int_{M_d} d^d x \sqrt{-\hat g}\left[
4\tilde a\hat R D_\mu n^\mu 
+ 2\tilde b\left(n_\mu n_\nu \hat R^{\mu\nu} 
D_\sigma n^\sigma + \hat R_{\mu\nu}D^\mu n^\nu \right) \right. \nn
&& \left. 
+ 8\tilde c n_\mu n_\nu \hat R^{\mu\tau\nu\sigma} D_\tau n_\sigma 
- {2 \over \tilde \kappa^2}D_\mu n^\mu \right] \nn
S_b^{(2)} &=& \int_{M_d} d^d x \sqrt{-\hat g}\left[\xi\tilde R
+ \eta \right] \ .
\eea
Here $\tilde R$ is the scalar curvature on the boundary and is defined via
 $\hat g_{ij}$ in (\ref{i}). 
In (\ref{Iiv}), $S_b^{(1)}$ is a generalization of 
(\ref{bdry2}) in Einstein gravity and corresponds to (\ref{HL2}). 
We also need to add $S_b^{(2)}$ (surface counterterm) which cancels the
divergence 
appearing in the limit of $\rho\rightarrow 0$ in the action. 
The divergence makes the energy-momentum tensor ill-defined. 
Note that $S_b^{(2)}$ is only given in terms of the boundary 
quantities, which do not affect the variational principle.   
If $\tilde \kappa = \kappa$, the term corresponding to 
$- {2 \over \tilde \kappa}D_\mu n^\mu$ is the term found in 
\cite{3}. If we choose $\tilde a = a+{2 \over d(d-1)}\hat c$, 
$\tilde b=-{4 \over d-1}\hat c$, $\tilde c=\hat c$ as in Weyl gravity
(see Eq.(\ref{viWa})), the boundary terms correspond to  
(\ref{HL1}). Since it is difficult to determine 
the coefficients $\tilde a$, $\tilde b$, $\tilde c$ 
and $\tilde\kappa$ by the variational principle in a closed 
form, we determine the coefficients $\tilde a$, $\tilde b$, 
$\tilde c$ and $\tilde\kappa$ and also $\eta$ and $\xi$ 
simultaneously from the condition of the cancellation of
the divergences. 
If $d\leq 4$, the terms in (\ref{Iiv})  
are enough to well-define the variational principle 
in the asymptotically AdS background. 
As we define these coefficients from the 
condition of cancellation of the divergence, not only 
$S_b^{(2)}$ will be surface counterterm but also part of 
surface counterterm will be included in $S_b^{(1)}$. 
It is difficult to divide $S_b^{(1)}$ to boundary term 
and surface counterterm (and such division
even looks unnatural as same terms with different
coefficients appear from both parts of surface terms) so we keep this term
mixed. The important thing will be that it leads to 
 finite gravitational action for asymptotically AdS spaces. 

In the coordinate choice (\ref{i}), $S_b$ in (\ref{Iiv}) has 
the following form:
\bea
\label{Iivb}
S_b&=&\lim_{\rho\rightarrow 0}
\int_{M_d} d^d x \sqrt{-\hat g}{\rho \over l}\left[
-4\tilde a\hat R \hat g_{ij} \left(\hat g^{ij}\right)'
- 2\tilde b\left({\hat R_\rho}^\rho \hat g_{ij} 
\left(\hat g^{ij}\right)' 
+ \hat R_{ij} \left(\hat g^{ij}\right)' \right) \right. \nn
&& \left. + 8\tilde c {\hat R^{\rho}}_{i\rho j} 
\left(\hat g^{ij}\right)' 
- {2 \over \tilde\kappa^2}\hat g_{ij} \left(\hat g^{ij}\right)'
+ {l \over \rho}\left(\xi\tilde R + \eta\right)\right]\ .
\eea
We now determine the coefficients $\tilde a$, $\tilde b$, 
$\tilde c$ and $\tilde\kappa$ and also $\eta$ and $\xi$ 
to cancel the divergent parts in the bulk action (\ref{vi}). 
The divergent parts, except the logarithmically divergent 
parts, in the bulk action is given by
\bea
\label{Si}
S&=&\int d^{d+1}x {l \rho^{-{d \over 2}-1} \over 2}
\sqrt{-g_{(0)}}\left[{4d^3 + 4d^2 \over l^4}a 
+{4d^2 \over l^4}b + {8d \over l^4}c - {2d \over l^2\kappa^2} 
\right. \nn
&& +\rho\left\{ R_{(0)}\left( -{2d(d+1) \over l^2}a - {2d \over l^2}b 
- {4 \over l^2}c + {1 \over \kappa^2}\right) \right. \nn
&& + g_{(0)}^{ij} g_{(1)ij}\left( {-4d^3 + 4d^2 + 8d \over 2l^4}a 
+ {-4d^2 + 8d \over 2l^4}b + {-4d + 8 \over l^4}c \right. \nn
&& \left.\left.\left. 
+ {2d - 4 \over 2l^2 \kappa^2}\right) \right\} + 
{\cal O}\left(\rho^2\right) \right] \ .
\eea
Here we expand the metric $g_{ij}$ as a power 
series with respect to $\rho$,
\be
\label{vii}
g_{ij}=g_{(0)ij}+\rho g_{(1)ij}+\rho^2 g_{(2)ij}+\cdots 
\ee
and the radius $l$ of AdS is found by solving the 
equation
\be
\label{Sii}
0={d^4 - 2d^3 - 3d^2 \over l^4}a + {d^3 - 3d^2 \over l^4}b 
+ {2d^2 - 6d \over l^4}c - {d^2 - d \over \kappa^2 l^2} 
- \Lambda\ ,
\ee
which follows from eqs. of motion. 
In Eq.(\ref{Si}), $\Lambda$ is deleted by using (\ref{Sii}). 

Especially the term proportional to $g_{(0)}^{ij} g_{(1)ij}$ 
in (\ref{Si}) would break the 
principle of the holography, where only $g_{(0)ij}$, which is 
the boundary value of the metric tensor $g_{ij}$, should give 
the boundary condition but $g_{(1)ij}$, which is the first order 
derivative of $g_{ij}$ with respect to $\rho$, should not do. 
If we treat $g_{(1)ij}$ as dynamical variable, we obtain the equation 
$g_{(0)ij}=0$ from the variation of $g_{(1)ij}$, which is 
inconsistent. 

For $d=4$ there also appear the following 
logarithmically divergent terms: 
\bea
\label{lndv}
S_{\rm ln}&=& \int d^{d+1}x {l \rho^{-{d \over 2}+1} \over 2}
\sqrt{-g_{(0)}}\left[a R_{(0)}^2 + b R_{(0)ij}R_{(0)}^{ij} 
+ c R_{(0)ijkl} R_{(0)}^{ijkl} \right. \nn
&& + R_{(0)}^{ij}g_{(1)ij}\left(l^2 K + {2d - 4 \over l^2}b \right)
- 3(d - 4) K g_{(0)}^{ij}g_{(2)ij} \nn
&& + R_{(0)} g_{(0)}^{ij}g_{(1)ij} \left( -{l^2 \over 2}K 
+ {4d - 4 \over l^2}a + {2 \over l^2}b 
+ {4 \over l^2}c \right)  \\
&& + g_{(0)}^{ij}g_{(0)}^{kl}g_{(1)ik}g_{(1)jl}
\left({3d -10 \over 2}K + {4d^2 - 16d + 16 \over 4l^4}b 
+ {4d -8 \over l^4}c\right) \nn
&& \left. + \left(g_{(0)}^{ij}g_{(1)ij}\right)^2\left(
-{3d -8 \over 4}K + {4d^2 - 8d + 4 \over l^4}a
+ {3d - 4 \over  l^4}b + {4 \over l^4}c \right)\right] \ .
\nonumber
\eea
Here
\be
\label{KK}
K \equiv {2d^2 + 2d \over l^4}a + {2d \over l^4}b 
+ {4 \over l^4}c - {1 \over l^2\kappa^2} \ .
\ee
The divergent part of the boundary action 
(\ref{Iiv}) is given by
\bea
\label{Siii}
\lefteqn{S_b=\lim_{\rho\rightarrow 0}\int d^d x 
{\rho^{1-{d \over 2}} \over l} \sqrt{-g_{(0)}}\left[
\left( {4d^3 + 4d^2 \over l^2}\tilde a + {4d^2 \over l^2}\tilde b 
+ {8d \over l^2}\tilde c - {2d \over \tilde\kappa^2} 
+ l\eta\right){1 \over \rho} \right.} \nn
&& + (-4d\tilde a - 2\tilde b + \xi l)R_{(0)} 
+ \left\{ {2d^3 -10d^2 + 4d \over l^2}\tilde a 
+ {2d^2 -8d +4 \over l^2}\tilde b + {4d - 8 \over l^2}\tilde c 
\right. \nn
&& \left.\left. + {-d + 2 \over \tilde \kappa^2} 
+ {l \over 2}\eta \right\}
g_{(0)}^{ij}g_{(1)ij} + {\cal O}(\rho)\right]\ .
\eea
 Comparing (\ref{Si}) and (\ref{Siii}), we find that the 
cancellation of the divergence requires
\bea
\label{Siv}
0&=&2d\tilde K l^2 + l\eta - 2Kl^2 \nn
0&=& -4d\tilde a - 2\tilde b + l\xi + {l^4 K \over d-2} \\
0&=& (2d+2)(-4d\tilde a - 2\tilde b) - (-d+2) \tilde K l^4 
+ {l^3\eta \over 2} + Kl^4\ .\nonumber
\eea
Here
\be
\label{Sv}
\tilde K \equiv {2d^2 + 2d \over l^4}\tilde a 
+ {2d \over l^4}\tilde b 
+ {4 \over l^4}\tilde c - {1 \over l^2\tilde\kappa^2} \ .
\ee
Eq.(\ref{Siv}) can be rewritten as follows:
\be
\label{Svi}
\tilde K=K-{d+1 \over l^4}A\ , \quad 
\eta = (2-2d)Kl - {2d(d+1) \over l^3}A \ ,
\quad \xi = - {l^3 \over d-2}K + {A \over l}\ .
\ee
Here 
\be
\label{Svii}
A\equiv 4d\tilde a + 2\tilde b\ .
\ee
The above equations seem not to be consistent with the result in 
\cite{HL}, which requires $\tilde K=K$ even if $A\neq 0$.  
As the number of the independent parameters, $\tilde a$, 
$\tilde b$, $\tilde c$, $\tilde \kappa$, $\xi$ and $\eta$, is six but 
the number of the equations in (\ref{Sv}) is three, there are three 
independent parameters. If we choose $\tilde a=\tilde b
=\tilde c=0$, we find
\be
\label{Sviii}
{1 \over \tilde\kappa^2}= -l^2K \ ,\quad 
\eta = (2-2d)Kl \ , \quad \xi = - {l^3 \over d-2}K \ .
\ee
Especially in case of pure Einstein theory with cosmological 
term ($a=b=c=0$), we find 
\be
\label{Sviiib}
{1 \over \tilde\kappa^2}= {1 \over \kappa^2} \ ,\quad 
\eta = {(2d-2) \over l\kappa^2} \ , \quad \xi 
= {l \over (d-2)\kappa^2} \ ,
\ee
This exactly corresponds to the counterterms in \cite{13}. 
When we choose $\tilde a=\tilde b
=\tilde c=0$, the energy-momentum tensor defined by $S_b^{(2)}$ 
has the following form
\be
\label{Six}
T_{ij}\equiv {1 \over \sqrt{-\tilde g}}{\delta S_b^{(2)} 
\over \delta \hat g^{ij}}
= \xi\left(\tilde R_{ij} - {1 \over 2}\tilde R \hat g_{ij}\right)
- {\eta \over 2}\hat g_{ij}\ .
\ee
The expression in (\ref{Six}) coincides with the full 
energy-momentum tensor in \cite{13} for the Einstein gravity. 

Let us consider the holographic trace anomaly making the choice  
$\tilde a=\tilde b =\tilde c=0$ in (\ref{Sviii}). 
First we consider 2d case, where we should drop $\xi$ term since 
it is finite. 
In order to control the logarithmically divergent terms in the 
bulk action $S$, we choose $d-2=\epsilon<0$. 
Then the total action $S+S_b$ has the following form
\bea
\label{2dTA}
\lefteqn{S+S_b = l^3 K \int_{M_{d+1}} 
d^{d+1}x {\rho^{1-{\epsilon \over 2}} 
\over 2}
\sqrt{-g_{(0)}} \left(-R_{(0)} - {\epsilon \over l^2}
g_{(0)}^{ij} g_{(1)ij}
+ {\cal O}\left(\rho\right)\right)} \nn
&=&l^3 K \int_{M_d} d^dx \left[{\rho_0^{1-{\epsilon \over 2}} 
\over \epsilon}
\sqrt{-g_{(0)}} \left(- R_{(0)} - {\epsilon \over l^2}
g_{(0)}^{ij} g_{(1)ij}\right) + {\cal O}
\left(\rho_0^{1-{\epsilon \over 2}}\right)\right] \ .
\eea
Here we replaced the integration over $M_{d+1}$ by 
\be
\label{replace}
\int_{M_{d+1}} 
d^{d+1}x \rightarrow \int_{M_d} d^dx \int_{\rho_0} d\rho 
\ee
with the finite (infrared) cutoff $\rho_0$, which is taken to be 
zero finally. 
We use the equation of the motion 
derived from  (\ref{lndv}) with respect to the variation over  
$g_{(1)ij}$ 
\bea
\label{g1i}
0&=&AR_{(0)}^{ij} + B g_{(0)}^{ij}R_{(0)} 
+ 2C g_{(0)}^{ik}g_{(0)}^{jl}g_{(1)kl}
+ 2D g_{(0)}^{ij}g_{(0)}^{kl}g_{(1)kl} \nn
A&\equiv&l^2 K + {2d-4 \over l^2}b \nn
B&\equiv&-{l^2 \over 2}K + {4d - 4 \over l^2}a + {2 \over l^2}b 
+ {4 \over l^2}c \nn
C&\equiv& {3d - 10 \over 2} K + {d^2 - 4d + 4 \over l^4}b 
+ {4d - 8 \over l^4}c \nn
D&\equiv& -{3d - 8 \over 4}K + {4d^2 - 8d + 4 \over l^4}a 
+ {3d-4 \over l^4}b + {4 \over l^4}c
\eea
and solve the equation with respect to $g_{(1)ij}$. Just for $d=2$, 
we obtain 
\be
\label{g1D2}
g_{(1)ij}=-{l^2 \over 4}R_{(0)ij}-{l^2 \over 8}R_{(0)}g_{(0)ij} 
=-{l^2 \over 4}R_{(0)}g_{(0)ij}\ .
\ee
(Note that $R_{(0)ij} = {1 \over 2}R_{(0)}g_{(0)ij}$ in 2 
dimensions.) Therefore we obtain 
\be
\label{g0g1D2}
g_{(0)}^{ij} g_{(1)ij}=-{l^2 \over 2}\left(1 
+ {\cal O}(\epsilon)\right)\ .
\ee
Since ${\delta \over \delta g_{(0)}^{ij}}\left(\int_{M_d}d^d x 
\sqrt{-g_{(0)}} R_{(0)}\right)=\sqrt{-g_{(0)}} \left(R_{(0)ij}
- {1 \over 2}g_{(0)ij}R_{(0)}\right)$, of course, we obtain
\be
\label{Eeq}
g_{(0)}^{ij}{\delta \over \delta g_{(0)}^{ij}}\left(\int_{M_d}
d^d x \sqrt{-g_{(0)}} R_{(0)}\right)=-{\epsilon \over 2}
\sqrt{-g_{(0)}} R_{(0)}\ .
\ee
Therefore only the leading term in (\ref{2dTA}) contributes to 
the trace anomaly and $g_{(0)}^{ij} g_{(1)ij}$ term, which gives 
the contribution of ${\cal O}(\epsilon)$, does not contribute.  
Then the trace anomaly for $d=2$ is given by 
\be
\label{Sx}
T=\lim_{\epsilon \rightarrow 0-}
{2\hat g_{(0)}^{ij} \over \sqrt{- \hat g_{(0)}}}{\delta (S+S_b) 
\over \delta \hat g_{(0)}^{ij}}
=l^3 K R_{(0)}\ , 
\ee
which is identical with the result found in \cite{NOha} where another
method was used. 
Note that there comes the contribution only from the bulk action. 
Although the equations of motion make the contribution vanish 
in the bulk but the additional contribution can appear from the surface 
due to the partial integration.

We now consider $d=4$ case. In this case we obtain, 
\be
\label{D4SS}
S+S_b=S_{\ln} +\ \mbox{finite terms}\ .
\ee
As in $d=2$ case, we use dimensional regularization putting 
$d=4+\epsilon < 0$. 
Eq.(\ref{g1i}) can be solved 
with respect to $g_{(1)ij}$ just for $d=4+\epsilon$ as follows 
\be
\label{g1D4}
g_{(1)ij}=-{l^2 \over 2}R_{(0)ij}
+ {l^2 \over 24}R_{(0)}g_{(0)ij} + {\cal O}(\epsilon)\ .
\ee
Substituting (\ref{g1D4}) into $S_{\ln}$ in (\ref{lndv}) 
and using the regularization in (\ref{replace}), we obtain 
\be
\label{D4ln}
S_{ln}={1 \over \epsilon}
\int_{M_d}d^d x \sqrt{-g_{(0)}}\left\{\left(
-{l^3 \over 8\kappa^2}+5al+bl\right)(G-F) 
+ {cl \over 2}(G+F) + {\cal O}(\epsilon)\right\}
\ee
Here we used the Gauss-Bonnet 
invariant $G$ and the square of the Weyl tensor $F$, 
which are given by
\bea
\label{GF} 
G&=&R_{(0)}^2 -4 R_{(0)ij}R_{(0)}^{ij} 
+ R_{(0)ijkl}R_{(0)}^{ijkl} \nn
F&=&{1 \over 3}R_{(0)}^2 -2 R_{(0)ij}R_{(0)}^{ij} 
+ R_{(0)ijkl}R_{(0)}^{ijkl} \ .
\eea
In (\ref{D4ln}), ${\cal O}(\epsilon)$ term contains 
$g_{(0)}^{ij} g_{(2)ij}$ and other terms are a sum 
of the square of the curvatures $R_{(0)}^2$, 
$R_{(0)ij}R_{(0)}^{ij}$ and $R_{(0)ijkl}R_{(0)}^{ijkl}$. 
If we use the equation of  motion, $g_{(0)}^{ij} g_{(2)ij}$ 
term can be given in terms of $g_{(0)ij}$. From the dimensional 
analysis, we can find that $g_{(0)}^{ij} g_{(2)ij}$ 
term is also given by a sum of the square of the curvatures and, 
if exists, the second order total derivatives of single curvature: 
$\Box_{(0)} R_{(0)}$ or $D_{(0)}^i D_{(0)}^j R_{(0)ij}$. The total 
derivative terms can be dropped in the action due to the volume 
integration over $M_d$. Since 
\bea
\label{TRsqr}
\lefteqn{g_{(0)}^{ij}{\delta \over \delta g_{(0)}^{ij}}
\left(\int_{M_d}d^d x 
\sqrt{-g_{(0)}} P_{(0)}\right)} \nn
&=&-{\epsilon \over 2}
\sqrt{-g_{(0)}} P_{(0)} + \ \left(\mbox{a sum of}\ \Box_{(0)} R_{(0)}\ 
\mbox{and}\ D_{(0)}^i D_{(0)}^j R_{(0)ij}\right) \nn
P_{(0)}&=&R_{(0)}^2,\ R_{(0)ij}R_{(0)}^{ij},\ 
R_{(0)ijkl}R_{(0)}^{ijkl}\ ,
\eea
and the second order total derivatives of single curvature: 
$\Box_{(0)} R_{(0)}$ and $D_{(0)}^i D_{(0)}^j R_{(0)ij}$, can 
be eliminated by a local counter terms if they appear in the 
trace anomaly, the ${\cal O}(\epsilon)$ terms in (\ref{D4ln}) do 
not contribute to the anomaly. We obtain the following expression 
for the trace anomaly for $d=4$
\be
\label{xvaa}
T=\left(-{l^3 \over 8\kappa^2}+5al+bl\right)(G-F) 
+ {cl \over 2}(G+F)\ .
\ee
The above result in (\ref{xvaa}) coincides with that in \cite{NOha}.
 From here it is easy to get the holographic trace anomaly for Weyl gravity.
Hence, we found the finite gravitational action (for asymptotically AdS
spaces) in HD gravity in three and
five dimensions by adding the local surface counterterm. This action 
correctly reproduces of holographic trace anomaly for dual (gauge) theory 
in d2 and d4. In principle, one can generalize all results for higher
dimensions,
say, d6, etc. With the growth of dimension, the technical problems become
more and more complicated as the number of structures in boundary term is
increasing.

\section{AdS black hole mass}

Now using the local surface counterterm prescription and the 
counterterm explicitly found in the previous section, we will find 
AdS black hole mass in HD gravity. 

When the metric of the 
reference spacetime (e.g. AdS) has the following form 
\be
\label{Mi}
ds^2 = f(r)dr^2- N^2(r)dt^2 
+ \sum_{i,j=1}^{d-1}\sigma_{ij}dx^i dx^j\ , 
\ee
the mass $M$ of the black hole like object is given by 
\be
\label{Mii}
M=\int d^{d-1}x \sqrt{\sigma} N \delta T_{tt}
\left(u^t\right)^2\ .
\ee
Here $\delta T_{tt}$ is the difference of the $(t,t)$ component 
of the energy-momentum tensor 
in the spacetime with black hole like object from that in the 
reference spacetime and $u^t$ is the $t$ component of the unit 
time-like vector normal to the hypersurface given 
by $t=$constant. 

First we consider the case of $c=0$ in (\ref{vi}), where the 
Schwarzschild anti-de Sitter black hole is the exact solution:
\bea
\label{Miii}
ds_{\mbox{\scriptsize S-AdS}}^2
&=&-\e^{-2\sigma} dt^2 + \e^{2\sigma}dr^2 + {r^2 \over l^2}
\sum_{i=1}^{d-1} \tilde g_{ij}dx^i dx^j \nn
\e^{-2\sigma}&=& {r^2 \over l^2}\left( {k r^{d-2} \over d-2} 
+ {\lambda^2 \over d(d-1)r^d} - \mu\right)\ .
\eea
Here $l$ is given by solving (\ref{Sii}) after putting $c=0$ 
and $\tilde g_{ij}$ is the metric of $d-1$ dimensional Einstein 
manifold, where the Ricci tensor $\tilde R_{ij}$ given 
by $\tilde g_{ij}$ is 
proportional to $\tilde g_{ij}$:
\be
\label{Miv}
\tilde R_{ij}=k\tilde g_{ij}\ .
\ee
We now consider the $d=4$ and $k=0$ case which appears in the 
throat limit of D3-brane. Then the $(t,t)$ component of the 
energy-momentum tensor in (\ref{Six}) has the following form:
\be
\label{Mv}
T_{tt}={\eta \over 2}\hat g_{tt}= {\eta \over 2}\left(-{r^2 \over l^2}
+ {r_0^4 \over l^2 r^2}\right)\ .
\ee
Note that the Ricci tensor $\tilde R_{ij}$ and the 
scalar curvature $\tilde R$  in (\ref{Six}) vanish since $k=0$, 
which tells the boundary space is flat. Then we find 
\be
\label{Mvi}
\delta T_{tt} = T_{tt} - T^{\mbox{\scriptsize AdS}}_{tt}
= {\eta r_0^4 \over 2l^2 r^2}\ ,\quad 
T^{\mbox{\scriptsize AdS}}_{tt} = - {\eta r^2 \over 2l^2}\ .
\ee
For the reference AdS metric, we find
\be
\label{Mvii}
\sqrt{\sigma}N={r^2 \over l^2}\ .
\ee
Then substituting (\ref{Mvi}) and (\ref{Mvii}) into 
(\ref{Mii}), the mass $M$ is 
\be
\label{Mviii}
M={\eta r_0^4 \over 2l^4}\int d^3 x\ , 
\ee
which agrees with the result obtained by using the 
standard prescription in \cite{NOst}. Especially, 
in the Einstein gravity case $a=b(=c)=0$, it reproduces the standard 
prescription  
result in \cite{GKT,GKP}:
\be
\label{Mix}
M={3 r_0^4 \over l^5\kappa^2}\int d^3 x\ .
\ee 

We now consider the Weyl gravity (\ref{viWb}). In this 
model, AdS is the exact solution and from (\ref{Sii}) we find 
its radius is identical with that of the Einstein gravity:
\be
\label{W1}
l^2=-{12 \over \Lambda \kappa^2}\ .
\ee
The Schwarzschild-AdS (S-AdS) solution is not, however, the exact 
solution of the system. Assuming $\hat c$ is small, we can 
solve S-AdS type solution perturbatively :
\bea
\label{W2}
\e^{-2\sigma}&=&{l^2 \over r^2}\left(1 - {r_0^4 \over r^4}
+ 2\epsilon {r_0^8 \over r^8}\right) \nn
\epsilon&\equiv& {\hat c \kappa^2 \over l^2}\ .
\eea
Here we make the coordinate choice as in (\ref{Miii}) :
\be
\label{MiiiW}
ds^2 =-\e^{-2\sigma} dt^2 + \e^{2\sigma}dr^2 + {r^2 \over l^2}
\sum_{i=1}^{d-1} \eta_{ij}dx^i dx^j \ .
\ee
As the difference of the asymptotic behavior from that of the S-AdS 
solution in the Einstein gravity appears only in the next-to-next 
order in ${1 \over r^4}$, we can evaluate the mass $M$ in the 
way similar to $c=0$ case in (\ref{Mviii})
\be
\label{MviiiW}
M={\eta r_0^4 \over 2l^4}\int d^3 x
= {3 r_0^4 \over l^5\kappa^2}\int d^3 x\ . 
\ee
In (\ref{MviiiW}), the $\epsilon$ ($\hat c$) correction does not 
appear and the result does not seem to agree with the standard prescription
 result 
in \cite{NOst}, where the mass (energy) is given by
\be
\label{W2b}
M={3 \left(\pi T\right)l^3 \over \kappa^2}
\left(1 + 18\epsilon\right)\int d^3 x \ .
\ee
Here the temperature $T$ is given by
\be
\label{W3}
\pi T = {r_0 \over l^2}\left(1 - {5 \over 2}\epsilon\right)\ .
\ee
Therefore we obtain 
\be
\label{W4}
M={3 r_0^4 \over l^5\kappa^2}\left(1 + 8\epsilon\right)
\int d^3 x \ .
\ee
In (\ref{W4}), there appear $\epsilon$ corrections in the 
next-to-leading order, which  conflict with (\ref{W2b}). 
This expresses the ambiguity in $R^2$ gravity in the choice of  
regularizations (standard prescription or local 
counterterm prescription) to make the action to be finite. In fact, we have 
chosen $\tilde a=\tilde b=\tilde c=0$ in Eq.(\ref{Sviii}) in 
order to fix the ambiguities in Eq.(\ref{Svi}). 
For general case when $\tilde a$, $\tilde b$ and $\tilde c$ do 
not vanish, $\eta$ in (\ref{Svi}) is shifted 
by $A$ in (\ref{Svii}). Then we obtain, 
instead of (\ref{MviiiW}), 
\be
\label{MviiiWb}
M={\eta r_0^4 \over 2l^4}\int d^3 x
= {3 r_0^4 \over l^5\kappa^2}\left(1 
- {20\kappa^2 \over 3l^2}A\right)\int d^3 x\ . 
\ee
Therefore the expression in (\ref{W4}) can be 
reproduced if we choose
\be
\label{WWi}
- {20\kappa^2 \over 3l^2}A=8\epsilon
\ee
or  using the expression for $\epsilon$ in (\ref{W2})
\be
\label{WWii}
A=-{6 \over 5}\hat c\ .
\ee
Then for general case with $c\neq 0$, it would be 
natural to choose
\be
\label{WWiii}
A=-{6 \over 5} c\ 
\ee
that is, in terms of $\eta$ and $\xi$ in (\ref{Svi})
for $d=4$, 
\be
\label{WWiv}
\eta=-6l\left({40a \over l^4} + {8\over l^4} 
 - {4c \over l^4} - {1 \over l^2\kappa^2}\right)\ ,
\quad 
\xi = -{l^3 \over 2}\left({40a \over l^4} + {8\over l^4} 
 + {8c \over 5l^4} - {1 \over l^2\kappa^2}\right)\ .
\ee
Even if we choose $A$ as in (\ref{WWiii}), of course, the 
previous results when $c=0$ do not change. 

As a check, we consider $a=b=0$ but $c\neq 0$ case. 
By the calculation similar to the previous 
case of the square of the Weyl tensor, we obtain
\be
\label{NNi}
M={\eta r_0^4 \over 2l^4}\int d^3 x
={3 r_0^4 \over l^4}\left(1 + {4\kappa^2 c \over l^2}\right)
\int d^3 x\ .
\ee
Note that from (\ref{WWiv}), $\eta$ is now given by
\be
\label{NNii}
\eta=-{6 \over l\kappa^2}\left( 1 
+ {4c\kappa^2 \over l^2}\right).
\ee
The result in (\ref{NNi}) agrees with the result 
obtained by using standard prescription in \cite{NOst}. 
In \cite{NOst}, the corresponding eq. is given by 
\bea
\label{NNiii}
&& M=E={3V_3\left(\pi T\right)^4 \over \kappa^2}\left(
1 + 12c\kappa^2\right) \nn
&& {1 \over \kappa^2}={N^2 \over 4\pi^2}\ ,\quad 
c\kappa^2 = {1 \over 16}\ ,\quad 
V_3=\int d^3x\ .
\eea
Here $l$ is chosen to be unity: $l=1$. 
Since the temperature $T$ is given by
\be
\label{NNiv}
\pi T =r_0\left(1 - 2c\kappa^2\right)\ ,
\ee
we find 
\be
\label{NNv}
M={3V_3 r_0^4 \over \kappa^2}\left(
1 + 4c\kappa^2\right)\ ,
\ee
which is identical with (\ref{NNi}) if we put $l=1$. 

Hence, we showed how for general HD gravity by fixing of 
unspecified parameters of local surface counterterm to make 
the complete correspondence between standard prescription
and local surface counterterm prescription (at least, 
in asymptotically AdS spaces). One can now repeat the similar calculation 
for another asymptotically AdS spaces, like Kerr-AdS one, etc.

\section{Discussion}

In summary, in this paper we constructed the local surface counterterm for 
HD gravity and for Weyl (stringy) gravity. Its coefficients are defined 
for general three- and five-dimensional HD gravity. Adding the local
surface counterterm to
bulk action of HD gravity makes the complete gravitational action to be
finite 
(at least for asymptotically AdS spaces). The gravitational energy-momentum 
tensor is getting also to be well-defined. As a by-product we obtain the
holographic 
conformal anomaly corresponding to d2 and d4 dual boundary quantum field
theory (via AdS/CFT correspondence). This result coincides with the
previous calculation 
made by different method.
The calculation of AdS black hole mass for HD gravity is made, using local
surface counterterm prescription. The corresponding mass (energy) for AdS 
space under consideration coincides with the standard prescription result.

Hence, our results indicate to universality of local surface counterterm
prescription developed originally for Einstein gravity. It is very
interesting to note that there are immediate directions on extension of our
results.
First of all, one could consider higher dimensional versions (say
seven-dimensional gravity) and to construct the corresponding local surface
counterterm and finite gravitational action in this case. This 
may be useful
in construction (via AdS/CFT correspondence) of dual six-dimensional 
quantum field theory (or, at least in consideration of next-to-leading 
corrections in such theory). Second, we limited our discussion to 
asymptotically AdS spaces. Clearly, one can generalize the local 
surface counterterm in HD gravity in the way indicated in ref.\cite{13} 
for Einstein gravity in order to make the gravitational action to be finite 
in asymptotically flat (or other non-compact) spaces. Third, having the
correct boundary term in HD gravity one may reconsider the circle of 
problems related with the role of higher derivative gravitational terms to
early Universe (for example, minisuperspace model). This will be
considered elsewhere.

\noindent
{\bf Acknowledgments.} 
The work by SDO has been partially supported by RFBR project N99-02-16617.

\end{document}